\documentclass[twocolumn,english,showpacs,superscriptaddress,aps,prl]{revtex4-1}
\usepackage[active]{srcltx}
\usepackage{color}
\usepackage{array}
\usepackage{textcomp}
\usepackage{multirow}
\usepackage{amstext}
\usepackage{graphicx}
\usepackage{bm}
\usepackage{pifont}
\usepackage{amsmath}
\usepackage{gensymb}
\usepackage{booktabs}
\makeatother
\usepackage{babel}
\begin{document}
\title{Floating Zone Growth of Sr Substituted Han Purple: Ba$_{0.9}$Sr$_{0.1}$CuSi$_{2}$O$_{6}$}
\author{Pascal Puphal}
\email{pascal.puphal@psi.ch}
\affiliation{Laboratory for Multiscale Materials Experiments, Paul Scherrer Institute,
5232 Villigen, Switzerland}
\author{Stephan Allenspach}
\affiliation{Neutrons and Muons Research Division, Paul Scherrer Institute, 5232 Villigen, Switzerland}
\affiliation{Department of Quantum Matter Physics, University of Geneva, 1205 Geneva, Switzerland}
\author{Christian R\"uegg}
\affiliation{Neutrons and Muons Research Division, Paul Scherrer Institute, 5232 Villigen, Switzerland}
\affiliation{Department of Quantum Matter Physics, University of Geneva, 1205 Geneva, Switzerland}
\author{Ekaterina Pomjakushina}
\affiliation{Laboratory for Multiscale Materials Experiments, Paul Scherrer Institute,
5232 Villigen, Switzerland}

\begin{abstract}
We present a route to grow single crystals of Ba$_{0.9}$Sr$_{0.1}$CuSi$_{2}$O$_{6}$
suitable for inelastic neutron studies via the floating zone technique.
Neutron single crystal diffraction was utilized to check their bulk quality and orientation. Finally, the high quality of the grown crystals was proven by X-ray diffraction and magnetic susceptibility.
\end{abstract}
\maketitle

\section{Introduction}

\begin{figure} [h]
\includegraphics[width=1\columnwidth]{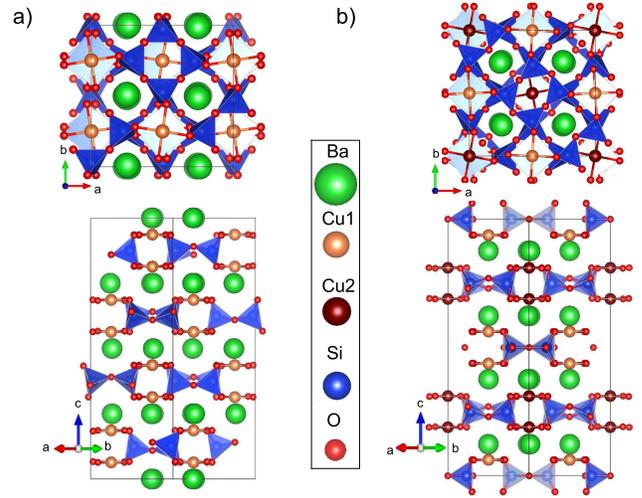}
\caption{(a) Room temperature $I4_{1}/acd$ structure of Han Purple shown along the $c$-axis and the (1 1 0) direction~\cite{Sparta2004}. (b) Same arrangement of the low temperature $Ibam$ structure of Han Purple~\cite{Sheptyakov2012}.}
\label{struc}
\end{figure}

Already in the Han Dynasty, the historically known compound Han Purple~\cite{Berke2007} BaCuSi$_{2}$O$_{6}$ was used as a purple coloring pigment in China. The~rather blue compound~\cite{Berke2007} was possibly created in ancient times with a lot of Cu$_{2}$O (red) inside, thus mixing to a purple pigment. Han Purple can be found in nature, e.g.,~Africa, as~the natural mineral Colinowensite~\cite{Rieck2015}. Rediscovered and reported upon in 1989~\cite{Finger1989}, it~attracted the interest of the physics community starting from 1997~\cite{Sasago1997}, when the rare arrangement of the Cu ions as pairs on a two dimensional square lattice (see Figure~\ref{struc}) was noticed to form dimers with a singlet ground-state and triplet excited states. As~the triplet state Zeeman split, in~high magnetic fields, a~two-dimensional Bose--Einstein condensate (BEC) of the bosonic triplet quasiparticles created a strong \mbox{interest~\cite{Jaime2004,Sebastian2006a}} in the compound. This interest was investigated after the discovery of an incommensurately modulated low temperature structure below 100\,K~\cite{Rueegg2007,Sheptyakov2012,Samulon2006} complicating the model and its physics. These~studies were performed on  BaCuSi$_{2}$O$_{6}$ crystals grown by two methods, namely, the floating zone (FZ) method in an oxygen flow~\cite{Jaime2004,Sparta2004} and from an oxygen spending lithiummetaborate flux~\cite{Sebastian2006,Sebastian2006a}, both with no detailed description of the growth conditions. Recently, we reported on the substitution series of (Ba,Sr)CuSi$_{2}$O$_{6}$, which stabilizes the tetragonal room temperature structure of BaCuSi$_{2}$O$_{6}$ down to lowest temperatures already with 5\% substititution~\cite{Puphal2016}. In~a following study, we could show the growth conditions of BaCuSi$_{2}$O$_{6}$ single crystals and its Sr-substituted variant~\cite{Well2016} by self melt growth in oxygen pressure. The~resulting crystals have a typical size of $2\times1\times0.5$\,mm$^3$. Here~we report on the floating zone growth of Ba$_{0.9}$Sr$_{0.1}$CuSi$_{2}$O$_{6}$, where large, high-quality single crystals are obtained for the first time, enabling future studies by, e.g.,~neutron~spectroscopy.

\section{Experimental~Details}

Thermogravimetric analysis was performed using a NETZSCH STA
409 analyzer. The~polycrystalline rods for the floating-zone
growth were pressed in a Powloka hydrostatic press. The~floating-zone growth was performed in a CSC FZ-1000-H-VI-VP-PC with a 300\,W halogen lamp (FZ1) and a SCIDRE HKZ equipped with a 5\,kW xenon lamp (FZ2). The~powder X-ray diffraction measurement was performed using a Bruker D8 Advance with a Cu cathode. Fluorescence spectra were recorded using the Orbis microXRF analyzer from EDAX. 
Neutron diffraction experiments were carried out on the MORPHEUS two-axis diffractometer at SINQ (PSI) at room temperature using a wavelength of $\lambda=5$\,\AA.
Magnetic susceptibility measurements were carried out in a range of 1.8–300\,K at 0.1\,T using
a quantum design physical property measurements system (PPMS). A~laboratory X-ray Laue equipped with CCD camera (Photonic Science) was used to orient the~samples.

\section{Synthesis}
Polycrystalline Ba$_{0.9}$Sr$_{0.1}$CuSi$_{2}$O$_{6}$ was prepared by sintering stoichiometric amounts of BaCO$_{3}$, SrCO$_{3}$, CuO, and~SiO$_{2}$. The~powder was ground and sintered in an aluminum
oxide crucible in air at 1028$\degree$C for 2 months, with several intermediate grindings to remove any early stage phases in the silicate formation \cite{Berke2007} as BaCu$_{2}$Si$_{2}$O$_{7}$. Its phase purity was checked with laboratory X-ray diffraction, proving to be of the $I4_{1}/acd$ structure (s.g. 142) shown in Figure \ref{struc} a. The~powder was then pressed into rods of a 7\,mm diameter by a hydrostatic press ($\sim$4000~bar) using rubber forms and subsequently annealed for 24 h in air at 1030$\degree$C. The~rod density was checked via dilatometry and found to be above 92\%. Finally, single crystals were grown using FZ1 as described~below.

\begin{figure} [h]
\includegraphics[width=1\columnwidth]{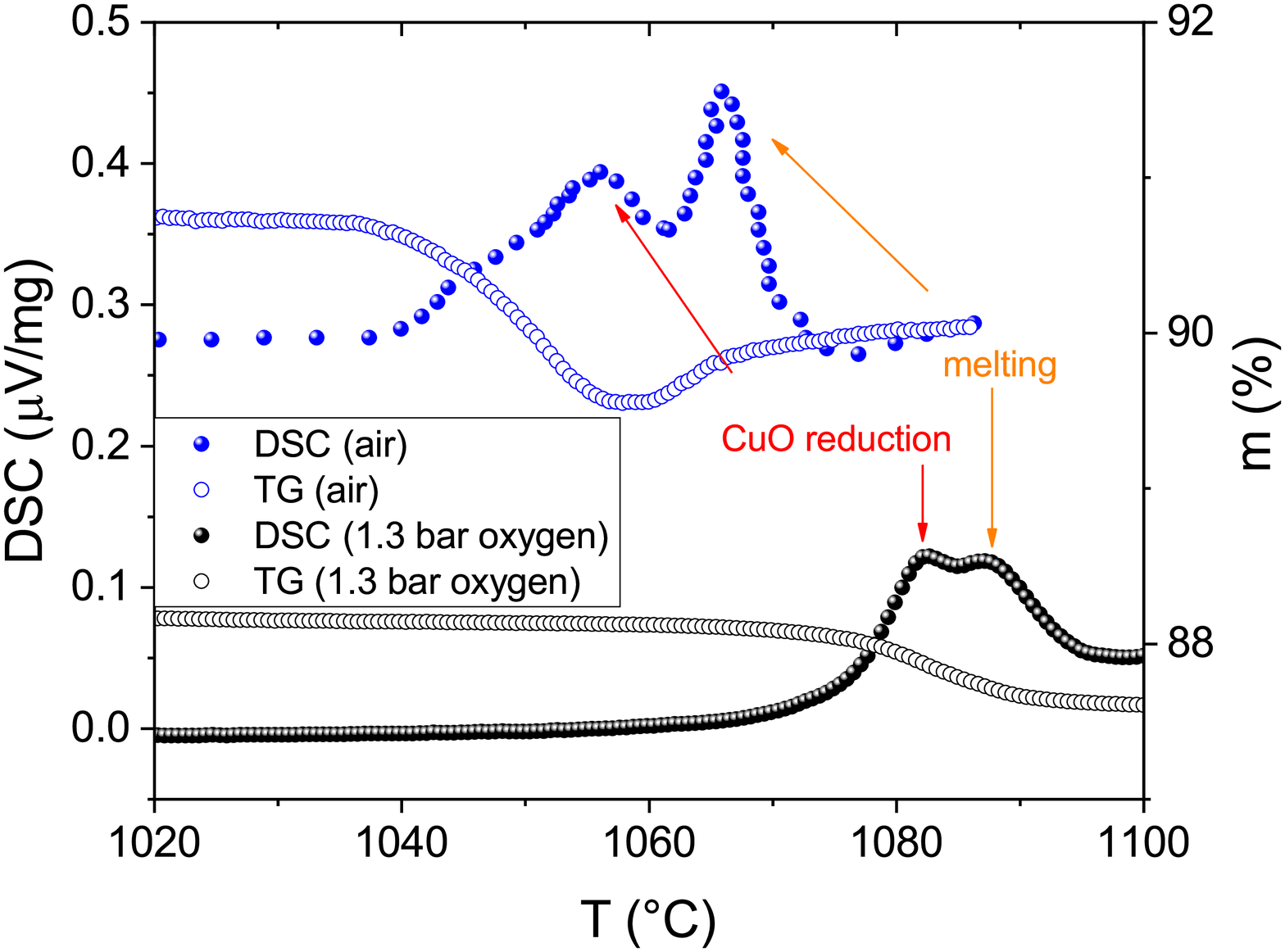} 
\caption{\textcolor{black}{\label{dta} Differential scanning calorimetry/thermogravimetric (DSC/TG) measurement of a stoichiometric BaCO$_{3}$, CuO, and~2 SiO$_{2}$ mixture upon heating in air and an oxygen partial pressure of 1.3 bar.}}
\end{figure}

We performed differential scanning calorimetry (DSC), including a thermogravimetric (TG) analysis, on the growth conditions of Ba$_{0.9}$Sr$_{0.1}$CuSi$_{2}$O$_{6}$ and observed the reduction of Cu$^{2+}$ to Cu$^{1+}$ while releasing oxygen (2CuO$\longrightarrow$ Cu$_{2}$O + $1\over{2}$O$_{2}$) monitored by a mass loss in the TG signal, followed directly by the melting of the compound. Afterwards, the~melting turns it to a viscous mass that glazes when cooled in air.
However, upon applying oxygen pressure, the decomposition is shifted up further than the melting temperature seen in a DSC experiment performed in air compared to one in oxygen flow with a partial pressure of 1.3\,bar~\cite{Well2016} (see Figure~\ref{dta}). Using a linear interpolation, the difference between decomposition and melting would meet at around 2\,bar. Thusm one would expect optimal growth conditions above this oxygen~pressure.

To obtain large (cm$^3$-size) single crystals of Ba$_{0.9}$Sr$_{0.1}$CuSi$_{2}$O$_{6}$, we utilized the floating zone growth method using two furnaces equiped with halogen lamps (FZ1) and a xenon lamp (FZ2). As~BaCuSi$_{2}$O$_{6}$ has a relative low melting point of around 1060~$\degree$C, which is slightly lowered by Sr substitution~\cite{Puphal2016}, a~low-power halogen lamp (FZ1) with a better focus can be applied. As~a first growth attempt, we followed the short report on the floating zone growth of BaCuSi$_{2}$O$_{6}$ from~\cite{Jaime2004}, and we used the same conditions for the substituted variant attempting a growth rate of 0.5\,mm/h in an oxygen flow of 200\,cc/min.
Stoichiometric seed and feed rods with a diameter of 7\,mm and a length of 7\,cm were used. With~these conditions, we~were unable to obtain a stable growth, as the reduction of CuO to Cu$_{2}$O led to bubble formations in the liquid zone, causing the rods to disconnect (see Table~\ref{growth} C1). 

\begin{figure} [h]
\includegraphics[width=1\columnwidth]{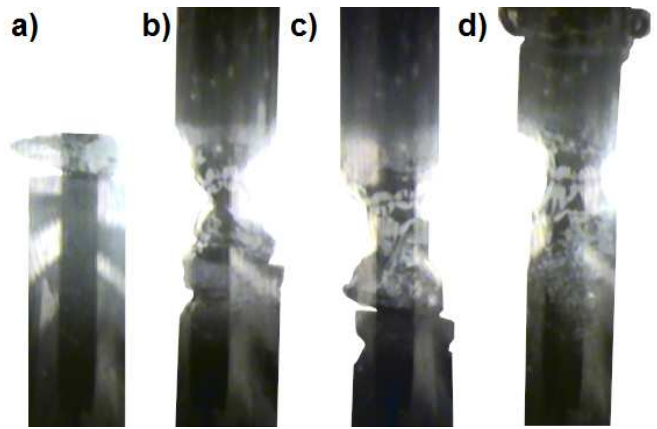} 
\caption{{\label{float} Images of the floating zone growth,
with (\textbf{a}) showing the glued crystal seed on the top of the
seed rod, (\textbf{b}) the start of the growth, and (\textbf{c},\textbf{d}) the evolution after several hours of growing. }}
\end{figure}

\begin{table*} [b]
\centering
\caption{\label{growth} Listed are the growth attempts of Ba$_{0.9}$Sr$_{0.1}$CuSi$_{2}$O$_{6}$ using stoichiometric seed and feed rods of 7\,cm length and 7\,mm diameter for FZ. Added as CF is the flux growth of Han Purple after~\cite{Sebastian2006}.}
\begin{tabular}{cccccccc}
\toprule & {\textbf{Furnace}} & {\textbf{Gas}} & \textbf{{p {[}bar{]}}} & \textbf{{Power [\%]} }& \textbf{{Rate {[}mm/h{]}}} & \textbf{{Comments}} & \textbf{{Crystallite Size}}\\
\midrule 
{C1} & {FZ1} & {O$_{2}$} & {0} & {53.7} & {-} & {could not start growth, immense bubbles} & {0}\\

{C2} & {FZ1} & {O$_{2}$} & {0.5--3} & {54--56.3} & {2-0.5} & {neck thinning and bubbles} & {$\mu$m}\\

{C3} & {FZ1} & {O$_{2}$} & {7} & {59} & {-} & {bubbles} & {0}\\

{C4} & {FZ2} & {O$_{2}$} & {30} & {16--18} & {1} & {flowing down of liquid, disconnection} & {0}\\

{C5} & {FZ2} & {O$_{2}$} & {100} & {22} & {2} & {repeated disconnection, phase seperation} & {0}\\

{C6} & {FZ1} & {O$_{2}$/Ar} & {4.4} & {56} & {1} & {stable growth conditions} & {mm}\\

{C7} & {FZ1} & {O$_{2}$/Ar} & {7} & {56.3} & {0.5} & {stable growth conditions} & {mm-cm}\\

{C8} & {FZ1} & {O$_{2}$/Ar} & {5.4} & {54.7} & {0.5} & {(seedcrystal) stable growth conditions} & {cm}\\

{CF} & {flux} & \multicolumn{4}{c}{{-{}-}} & {2:1 LiBO$_{2}$, $1000\degree$ C slow cooling to $875~\degree$ C} & {mm}\\
\bottomrule 
\end{tabular}
\end{table*}

\subsection{Atmosphere}
Surprisingly, an increase of pressure with attempts at 0.5, 1, 2.5, 3, 4, 5, and 7\,bar led to similar problems and could not stabilize the melt and suppress the bubble formation (C2,3). In~all cases, when changing the lamp power by a few percentage points, the~melt was either too viscous at low lamp power or~the temperature was too high, leading to a decomposition. This implies that the decomposition and melting point are still overlapping, as seen in the DSC curve. 
As CuO can only be grown at elevated oxygen pressures~\cite{Prabhakaran2003,TAKAYUKI1998}, we wanted to attempt even higher pressures to ensure a clear separation of the CuO reduction and melting point following an extrapolation of Figure~\ref{dta}. Even upon application of 100\,bar of oxygen pressure (using FZ2), we  were still unable to stabilize the liquid zone (C4). Repeated disconnection complicated the growth attempt, and afterwards, we could still find orange parts of reduced copper oxide on crushed pieces of the obtained crystal. 
Thus, in~order to find stable growth conditions with FZ1, the~experience of applying an oxygen--argon mixture in the comparable compound of SrCu$_{2}$(BO$_{3}$)$_{2}$ \cite{Zayed2014, Dabkowska2007} was used. Varying the pressure and oxygen/argon mixture leads to a liquidification of the melt and a breaking of the bubbles into smaller, not visible, ones with increasing argon content, which stabilizes the growth conditions. We found that the concentration of argon has to exceed the oxygen one for this effect. Nevertheless, in~a cross section of obtained multigrain crystals, there is still some orange Cu$_{2}$O phase present, proving the continuation of a decomposition at any attempted pressure and mixture of gases. Without~the possibility of a full suppression of the CuO reduction, we studied the growth in stable conditions using four 300\,W lamps operated at 54.5\% with 5\,bar pressure obtained with 0.4\,L/min argon and 0.1\,L/min oxygen gas-flow: Only in the case of several simultaneously growing grains is the Cu$_{2}$O impurity incorporated in between the grains, while with a single grain, the impurity is pushed up with the liquid zone until the very last part of the melt. Thus, the orange Cu$_{2}$O is kept in the molten zone without any influence on the single crystal~formation.
\subsection{Growth~Speed}
\begin{figure}
\includegraphics[width=1\columnwidth]{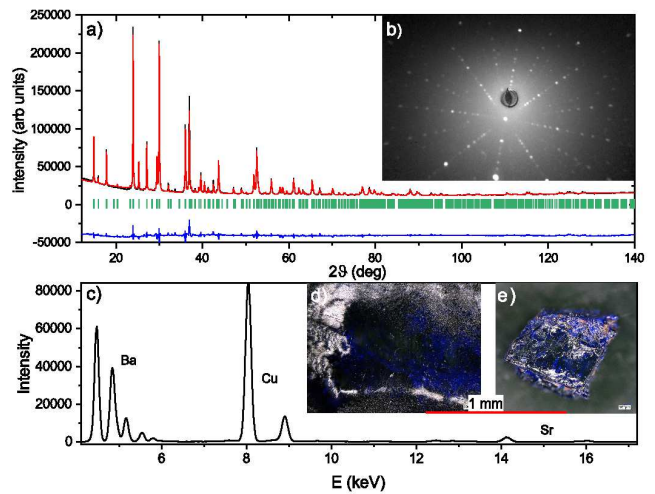} 
\caption{{\label{x-ray} (\textbf{a}) Rietveld refinement of the crystal structure parameters of
Ba$_{0.9}$Sr$_{0.1}$CuSi$_{2}$O$_{6}$ crushed single crystals, based on laboratory X-ray powder diffraction data at 295\,K. The~rows of ticks in the middle correspond to the calculated diffraction peak positions of the $I4_1/acd$ structure. (\textbf{b}) Laue image on the shiny side of the single crystal shown in Figure~\ref{neut}b revealing the $c$-axis of the tetragonal system. (\textbf{c})~MicroXRF spectra of the Ba$_{0.9}$Sr$_{0.1}$CuSi$_{2}$O$_{6}$ single crystal C8. (\textbf{d}) Magnified image of a section from the FZ crystal C8 (\textbf{e}) Broken piece of an FZ grown crystal solely in oxygen atmosphere C2.}}
\end{figure}

\begin{figure} [h]
\includegraphics[width=1\columnwidth]{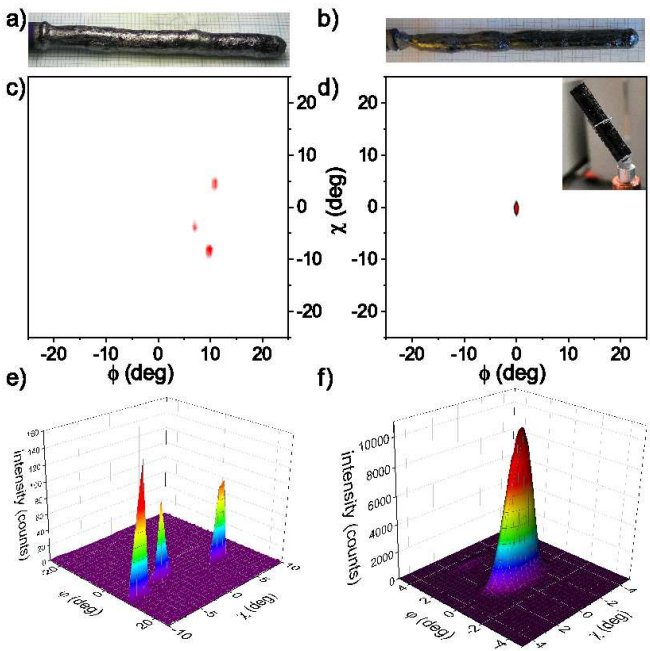} 
\caption{{\label{neut} Comparison of crystal C6 grown with 1\,mm/h (\textbf{a},\textbf{c},\textbf{e}) and C8 0.5\,mm/h (\textbf{b},\textbf{d},\textbf{f}). Image of the floating zone grown crystal is given in (a,b) and then their neutron scattering scans of the (0 0 4) reflection plotted versus two rotation angles. First, a contour plot in (c,d) shows a larger area displaying the amount of grains and second, a 3D surface plot in (e,f) reveals the detailed shape.}}
\end{figure}

Ba$_{0.9}$Sr$_{0.1}$CuSi$_{2}$O$_{6}$ crystals grow quickly along the $a$-direction, while the $c$-direction develops much slower, leading to growth steps and, thus, faceted structures (terraces) of the crystal flakes grown by direct melt~\cite{Well2016}. This leads to the a-axis developing as the growth direction in FZ experiments, while the $c$-axis points perpendicular to the round surface. With~growth rates of 1\,mm/h, several grains develop with a tilt in the $ab$-plane limited by the c-direction growth rate (C6). We thus reduced the growth rate to 0.5\,mm/h (C7) and additionally started to grow with a single crystal as a seed (C8) oriented with the c axis pointing along the growth direction (see Figure~\ref{float}a,b) to prevent the formation of the misaligned grains. The~seed-crystal was glued to the seed-rod by GEvarnish and then molten to it with the optical furnace by quickly melting the tip of the seed-rod, where the organic glue is fully burned away. Then the growth was started by melting the feed only and moving seed and feed together. Meanwhile, the~forced growth orientation along the \textit{c}-direction was not successful, as throughout the growth the growing crystal went back to the preferred orientation, the~prevention of additional grains was successful in the entire grown crystal. Even~with the chosen method of floating zone growth, the usually round crystal shows a shiny facet on both sides perpendicular to the \textit{c} direction (see {Figure}~\ref{x-ray}d). In this case, as~there was a reorientation, which was not completed after the end of the growth, there is an angle of 30$\degree$~between the growth direction and the \textit{a}-axis (see inset of {Figure} \ref{neut}d).

\unskip

\section{Characterization}

Ba$_{0.9}$Sr$_{0.1}$CuSi$_{2}$O$_{6}$ crystals cleave well perpendicular to the $c$-direction, leaving shiny \textit{ab}-planes enabling an easy orientation when breaking off the last and first part of the growth. Obtained Ba$_{0.9}$Sr$_{0.1}$CuSi$_{2}$O$_{6}$ crystals were analyzed in a neutron scattering experiment using the MORPHEUS 2-axis diffractometer at SINQ (PSI) at room temperature with a wavelength of 5\,\AA \, (see Figure~\ref{neut}c--f). We~scanned one reflection by fixing the detector to, e.g.,~$2\theta=$53.34$\degree$ \, for~the  (0 0 4) reflection shown in Figure~\ref{neut} and measured while rotating the crystal to search for additional grains. As~discussed above, this neutron diffraction measurement revealed that earlier growth attempts with 1 mm/h gave rise to three grains developing rather equally throughout the rod. They are slightly tilted in the \textit{ab}-plane in respect of each other but~share a similar \textit{c}-axis orientation perpendicular to the growth direction (see Figure~\ref{neut}c,e). 

In the seeded 0.5\,mm/h growth (C8), a full 180$\degree$ \,rotation scan of the (0 0 4) reflection showed only one grain in the entire piece, giving a sizeable single crystal of several grams and 4\,cm length (see inset of Figure~\ref{neut}d). We found a broadening of the reflection along one angle (see Figure~\ref{neut}f) due to the change of the growth direction along the crystal length.

\begin{figure} [h]
\begin{centering}
\includegraphics[width=1\columnwidth]{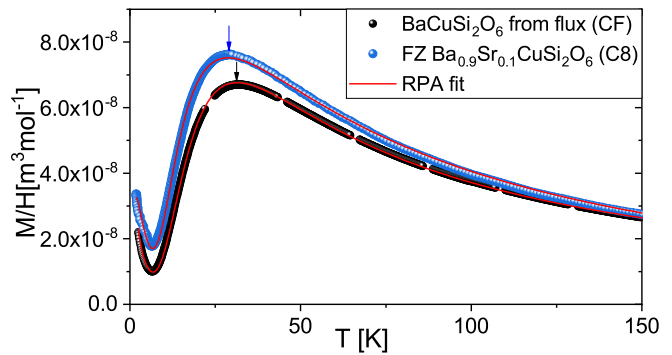} 
\par\end{centering}
\caption{\textcolor{black}{\label{susc} Temperature-dependent magnetic susceptibility
of the two single crystals C8 and CF in the range of 1.8-300\,K measured in a field
of 0.1\,T along the \textit{a}-direction. A~slight offset was chosen for visibility.}}
\end{figure}

We reproduced the flux growth for non substituted (as the flux reacts with Sr) Han Purple reported in~\cite{Sebastian2006} and compared the single crystal quality of both samples. All flux grown crystals as well as crystals prepared with oxygen pressure in the way described in~\cite{Well2016} and by the floating zone growth reported here can be obtained with equal quality (in the sense of magnetic impurities). 
In both ways, by direct melting with oxygen pressure and with flux, the magnetic impurity amount can range from 3 to 25\%. The~magnetic impurities can be analyzed in this system  by fitting the Curie tail arising from free spin-1/2 levels (e.g., paramagnetic BaCuSi$_{4}$O$_{10}$ ) at low temperatures, which is entirely
dominated by impurities. In~Figure~\ref{susc}, we show a magnetic susceptibility measurement comparing high-quality crystals of BaCuSi$_{2}$O$_{6}$ grown with the flux technique to the Ba$_{0.9}$Sr$_{0.1}$CuSi$_{2}$O$_{6}$ single crystal grown with the floating zone technique. Both have a similar impurity amount of less than 1\% magnetic impurities. The~shift of the maximum, due to a smaller unit cell, for~Sr substitution is apparent (see arrows in Figure~\ref{susc}). For~the antiferromagnetic intradimer coupling parameter, a random phase approximation (RPA) fit yields $J_{float}=45.9(1)$\,K compared to $J_{flux}=51.3(3)$\,K. This hints at a full incorporation of 10\% Sr following $J_{Sr}\approx(50-40x_{Sr})$\,K, in agreement with the microXRF results of 9(2)\% (see Figure~\ref{x-ray}c) and lattice constants of $a\approx 9.96206(9)\,\AA$, $c\approx22.2871(2)\,\AA$ (obtained by the Rietveld refinement shown in Figure~\ref{x-ray}a), matching the published results from a powder sample~\cite{Puphal2016}.
The nonmagnetic Cu$_{2}$O can be observed optically as an orange impurity and changes the color from blue to purple. With~its sharp contrast to blue, this impurity can be seen on cleaved surfaces in the microscope. In~Figure~\ref{x-ray}d, we show first a magnification of the facet on the side of the floating zone crystal C8 and, second, a broken piece of an FZ growth attempt in oxygen pressure C2. In~the second case, one can clearly see orange parts of reduced copper oxide, which is not present in the final growth. The~Laue image in Figure~\ref{x-ray}b was taken on the surface of C8, i.e.,~the facet of (d), %Does this refer to a figure? If so, please follow figure format, "Figure Xd".
proving the $c$-direction orientation and crystal~quality.

\section{Summary}

Via the optical floating zone growth of Ba$_{0.9}$Sr$_{0.1}$CuSi$_{2}$O$_{6}$,
we obtained large single crystals suitable for further neutron studies.
With an 80\%--20\% argon--oxygen mixture, stable conditions could be
created at 5\,bar of pressure. With~these conditions, using four 300\,W
lamps at 54.5\%, a single crystal is obtained, growing with a slow rate
of 0.5\,mm/h. The~use of a seed crystal proved crucial to get a large single grain at this growth rate. By~neutron single-crystal diffraction, we proved that a single grain is obtained, with~the same crystal quality as flux grown BaCuSi$_{2}$O$_{6}$.
\vspace{6pt}

\acknowledgments{This work is partly based on experiments performed at the
Swiss spallation neutron source SINQ, Paul Scherrer Institute,
Villigen, Switzerland.
The measurements were carried out on the PPMS/MPMS devices of the Laboratory for Multiscale Materials Experiments, Paul Scherrer Institute, Villigen, Switzerland.
The authors would like to acknowledge the Swiss National Science Foundations
(SNSF R\textquoteright Equip, Grant No. 206021\_163997 and Grant No.~206021\_139082) and matching funds from Paul Scherrer Institute for
purchasing the SCIDRE HKZ---high pressure high-temperature optical floating
zone furnace and the MPMS. 
}
\bibliographystyle{apsrev4-1}
\addcontentsline{toc}{section}{\refname}\bibliography{HP}
\end{document}